\documentclass[sigconf]{acmart}

\AtBeginDocument{%
  }

\setcopyright{acmcopyright}
\copyrightyear{2018}
\acmYear{2018}
\acmDOI{XXXXXXX.XXXXXXX}

\acmConference[Conference acronym 'XX]{Make sure to enter the correct
  conference title from your rights confirmation emai}{June 03--05,
  2018}{Woodstock, NY}
\acmPrice{15.00}
\acmISBN{978-1-4503-XXXX-X/18/06}

\graphicspath{{./figs/}}

\usepackage{graphicx}
\usepackage{subfigure}
\usepackage{pifont}
\usepackage{threeparttable,booktabs,multirow,lscape}
\usepackage{adjustbox}
\usepackage{enumerate}
\usepackage{listings}
\usepackage{float}
\usepackage{xspace}
\usepackage{soul}

\usepackage{flafter}
\usepackage{stfloats}

\usepackage{flushend}

\emergencystretch=\maxdimen
\newcommand{\todo}[1]{\textcolor{purple}{[todo: #1]}}

\newcommand{\etal}{\emph{et al.}\xspace}
\newcommand{\etc}{\emph{etc}\xspace}
\newcommand{\ie}{\emph{i.e.}\xspace}
\newcommand{\eg}{\emph{e.g.}\xspace}
\newcommand{\aka}{\emph{a.k.a.}\xspace}

\begin{document}

\title{Measurement of the Usage of Web Clips in Underground Economy}

\author{Qinyu Hu}
\authornotemark[1]
\affiliation{%
  \institution{Shanghai University of Engineering Science}
  \city{Shanghai}
  \country{China}}
\email{hqy@sues.edu.cn}

\author{Songyang Wu}
\authornote{Both authors contributed equally to this research.}
\affiliation{%
  \institution{The Third Research Institute of The Ministry of Public Security}
  \streetaddress{1 Th{\o}rv{\"a}ld Circle}
  \city{Shanghai}
  \country{China}}
\email{wusongyang@stars.org.cn}

\author{Wenqi Sun}
\authornote{Corresponding authors}
\affiliation{%
 \institution{The Third Research Institute of The Ministry of Public Security}
  \city{Shanghai}
  \country{China}}
\email{sunwenqi@stars.org.cn}

\author{Zhushou Tang}
\affiliation{%
  \institution{QI-ANXIN Technology Group Inc.}
  \streetaddress{30 Shuangqing Rd}
  \city{Shanghai}
  \country{China}}
\email{ellison.tang@gmail.com}

\author{Chaofan Chen}
\affiliation{%
 \institution{QI-ANXIN Technology Group Inc.}
  \city{Shanghai}
  \country{China}}
\email{darenfy@gmail.com}

\author{Zhiguo	Ding}
\affiliation{%
 \institution{The Third Research Institute of The Ministry of Public Security}
  \city{Shanghai}
  \country{China}}
\email{dingzhiguo@stars.org.cn}

\author{Xiaomei Zhang}
\authornotemark[2]
\affiliation{%
 \institution{Shanghai University of Engineering Science}
  \city{Shanghai}
  \country{China}}
\email{xmzhang@sues.edu.cn}


\renewcommand{\shortauthors}{Qinyu Hu, Songyang Wu et al.}



\begin{abstract}
In this paper, we study the ecosystem of the abused Web Clips in underground economy. Through this study, we find the Web Clips is wildly used by perpetrators to penetrate iOS devices to gain profit. This work starts with 1,800 user complaint documents about cyber crimes over Web Clips. We firstly look into the ecosystem of abused Web Clips and point out the main participants and workflow. In addition, what is the Web Clips used for is demystified. Then the main participants, including creators, distributors, and operators are deeply studied based on our dataset. We try to reveal the prominent features of the illicit Web Clips and give some mitigation measures.

Analysis reveals that 1) SSL certificate is overwhelmingly preferred for signing Web Clips instances compared with certificate issued by Apple. The wildly used SSL certificates can be aggregated into a limited group. 2) The content of the abused Web Clips falls into a few categories, `Gambling', `Fraud', and `Pornography' are among the top categories. 3) Instant messenger (IM) and live streaming platform are the most popular medium to trick victims into deploying the Web Clips. 4) The Web Clips are operated by a small amount of perpetrators, and the perpetrators tend to evade detection by taking technical approach, such as registering domain names through oversea 
domain name service provider, preferring easy-to-acquire new gTLD (global Top Level Domain), and deploying anti-crawler tricks. 

Our study gives hints on investigation of cyber crime over Web Clips, we hope that this work can help stakeholders to stay ahead of the threat.
\end{abstract}


\maketitle


\section{Introduction}\label{sec:Introduction}
Cyber crime is a pervasive and costly global issue. 
According to the investigation of Internet Crime Complaint Center~\cite{ic3_report}, the loss caused by cyber crime is increasing every year. In 2021, the reported losses of the cyber crime exceed \$6.9 billion, with an over 50\% increase than last year. Other report from China indicates that the law enforcement has identified and frozen over 300 billion RMB of the domestic transactions of cyber crime in 2021~\cite{Fraud_China}.

There is a wide range of vectors used by perpetrators in the criminal activity, including emails, websites, or advertisements.
In order to protect end-user against the cyber crime, lots of studies~\cite{almomani2013survey, srivastava2008credit, mansfield2014dark} have paid attention to such vectors that facilitate cyber crimes. Recently, researchers dedicate to vetting the mobile applications with the ever-evolving threats to mobile users~\cite{chen2021lifting, honganalyzing}. 
Among these research, a plethora of works target the APK file on Android, whereas few has paid attention to APPs on iOS~\cite{lee2019understanding}. 

Apple has a good track record of protecting end-user from malicious APPs. To guarantee the safety of the APP, Apple vets all APPs~\cite{apple_guide} before they're published on the App Store. Moreover, Apple does not allow for sideloading APPs, requiring that any APP installed on the device is distributed through the iTunes App Store.  In order to breach such rule to reach iOS users, the corner content distribution channels (\eg, iMessage~\cite{seriesmobile}, TestFlight~\cite{test_flight} or Web Clips~\cite{web_clips}) are utilized by miscreants. Among these corner channels, Web Clips, which are used to streamline the process of setting up a large number of iOS devices, are abused for illicit content distribution.

This paper aims to answer the four research questions to demystify the Web Clips in underground economy, which are beneficial for cyber investigation and new ways to prevent cyber crimes. 
\begin{itemize}
    \item [\textbf{RQ1.}]Who create Web Clips and by what? 
    \item [\textbf{RQ2.}]What is the purpose that the Web Clips are used for?
    \item [\textbf{RQ3.}]How are the illicit Web Clips distributed? 
    \item [\textbf{RQ4.}]What are the characteristics of operators in the Web Clips based criminal activities? 
\end{itemize}

In order to demystify the ecosystem of the abused Web Clips and pinpoint the participants of the underground economy, we study the Web Clips dataset in a pipeline and give insight on a variety of aspects, including the creator, distributor, and operator of the Web Clips. Our findings include: 1) SSL certificate is overwhelmingly preferred for signing Web Clips instances compared with certificate issued by Apple. The wildly used SSL certificates can be aggregated into a limited group, which can be  correlated with online Web Clips generating service. Since the online generators are so widely used in underground economy, we highly recommend that more regulatory measures should be taken. 2) The content of the abused Web Clips falls into a few categories, `Gambling', `Fraud', and `Pornography' are among the top categories. These are also the most pervasive cyber crimes. 3) Instant messenger and live streaming platform are the most popular medium to trick victims into deploying the Web clips. About 66\% illicit Web Clips are distributed through instant messenger and live streaming.  4) The Web Clips are operated by a small amount of perpetrators since the Web Clips can be aggregated by using a variety of features. And the perpetrators tend to escape detection by taking technical approach, such as registering domain names through oversea organization, preferring easy-to-acquire new gTLD, and deploying anti-crawler tricks. 

In this paper, we make the first systematic study on the Web Clips ecosystem especially being abused in illicit activities. Firstly, we introduce research background, the used dataset of this work and the Web Clips ecosystem overview (see section~\ref{sec:Background}). Creators perform the generation of Web Clips using local tools or online Web Clips generators (see section~\ref{sec:creator}). Then the purpose of the illicit Web Clips are analyzed (see section~\ref{sec:purpose}). Distributors spread Web Clips over various medium (see section~\ref{sec:distributor}). Operators deploy and operate the Web Clips and related websites (see section~\ref{sec:operator}). At last, based on the research findings of Web Clips ecosystem, mitigation measures are recommended (see section~\ref{sec:mitigation}).

\section{Background}\label{sec:Background}
As a daily-use device for storing personal information and sensitive data, smartphones are a perfect target for cyber crimes nowadays. 
Stakeholders, including app stores or mobile operating systems, have made tremendous effort to protect the end-users from falling prey to cyber crimes.
The safety net erected by stakeholders drives malware/scamware/ransomware to find new ways to reach smartphone systems. 
Among these ways, Web Clips, which are used for configuring iOS devices, are abused to break the confinement.  

Similar to the android APP generated by a generator~\cite{oltrogge2018rise}, Web Clips of iOS can be generated for developers with little or no software engineering background (\aka, citizen developer) by using generators as well. 
Security concern of android APP generated by a generator has attracted ample researches~\cite{oltrogge2018rise, chen2021lifting, honganalyzing}, however, the Web Clips are still lack of specific study and understandings so far.



In this section, we start with the introduction of Web Clips, then the initial dataset being used is described, furthermore we give a glimpse at the ecosystem of abused Web Clips.
\subsection{Apple Web Clips} \label{sec:web_clip_intro}

\begin{figure}
    \begin{minipage}[t]{0.49\linewidth}
        \subfigure[Widget added by using Web Clips]{
        \label{fig:widget}
        \includegraphics[width=1.4in]{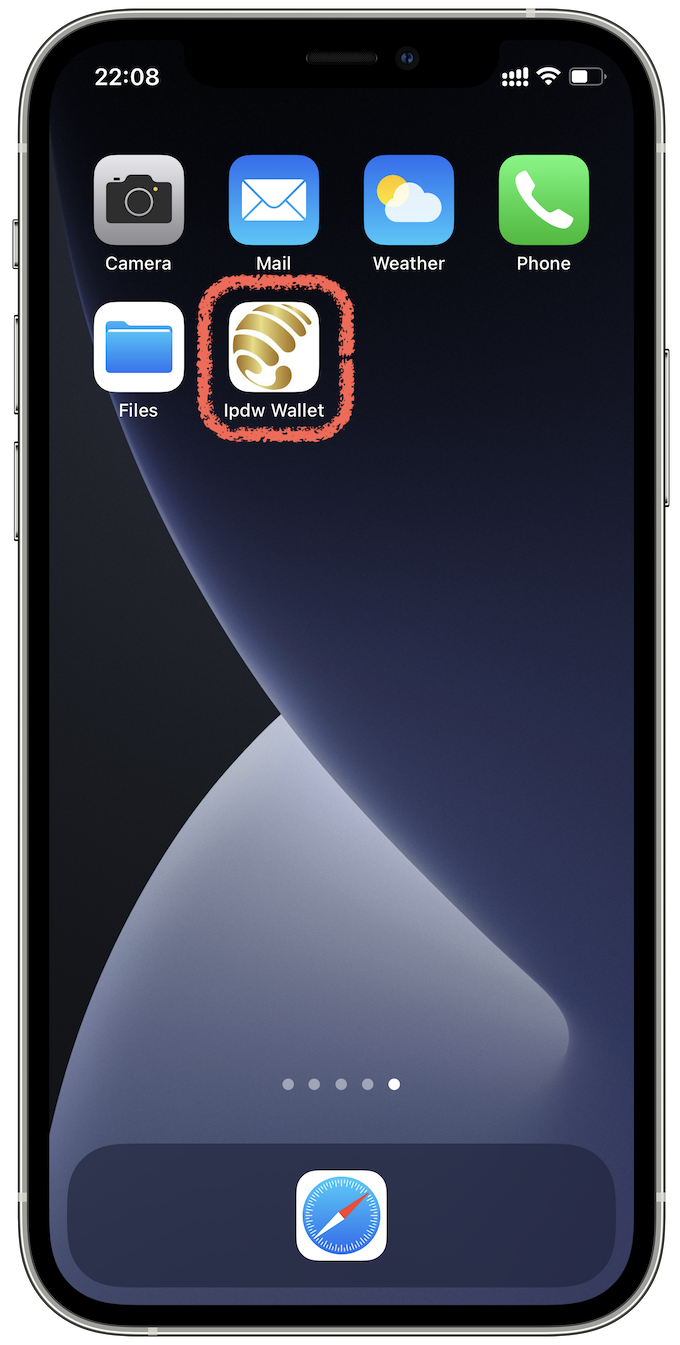}}
    \end{minipage}
    \begin{minipage}[t]{0.49\linewidth}
        \subfigure[View of the widget (no navigation bar)]{
        \label{fig:widget_view}
        \includegraphics[width=1.4in]{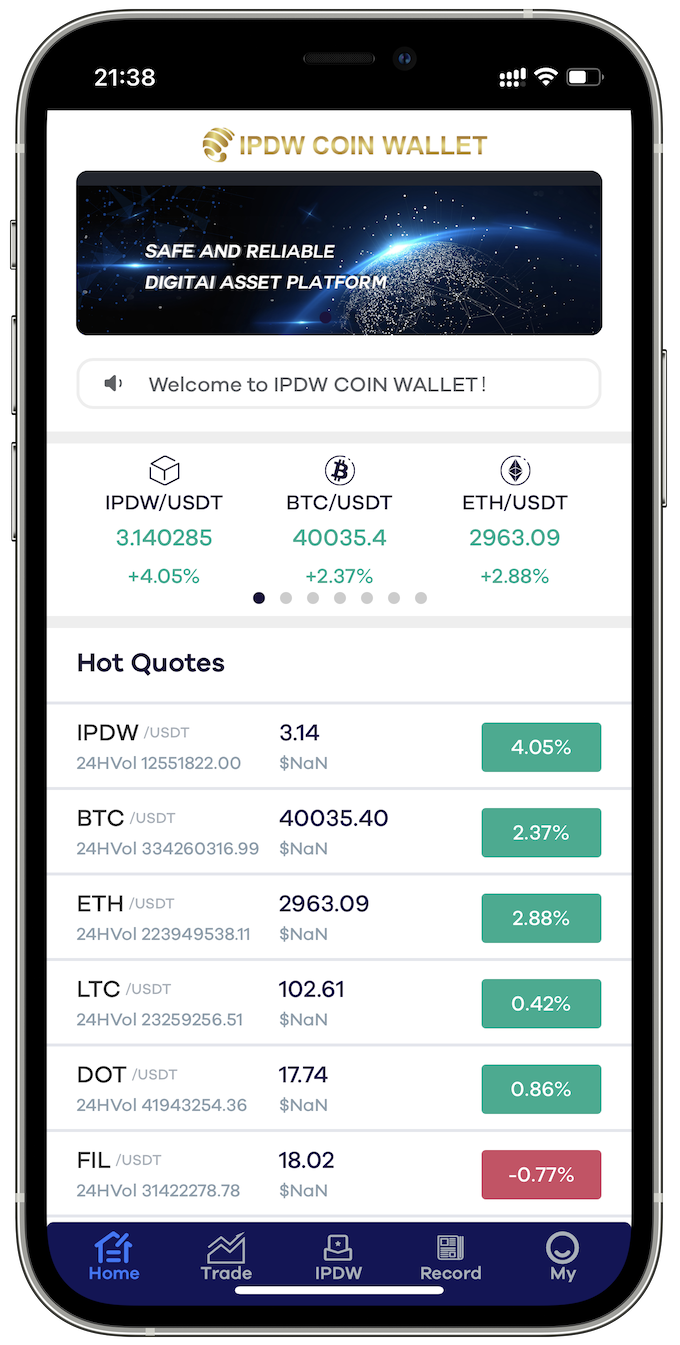}}
    \end{minipage}
    
    \vspace{-0.2cm}
    \caption{The widget looks like an app intuitively.}
    \label{fig:web_clip_example}
    \vspace{-0.2cm}
\end{figure}

Apple allows to configure a large scale of enrolled iOS-based devices to the customized setting via \emph{configuration profiles}~\cite{ConfiguringDev, OTAProfile, ConfigurationProfileReference} in mobile device management~\cite{web_clips_2} scheme. The configuration profiles contain a rich set of settings for configuring email, network, \etc, of iOS devices. Among these settings, the Web Clips (The configuration profiles containing \emph{Web Clips setting} are referred to as \emph{Web Clips} throughout this paper for brevity) allow to add a widget to the Home Screen of the user's device. The widget provides fast access to website links. After the Web Clip (the file end with .mobileconfig) has been delivered and deployed, it looks like an APP is installed, users can tap the widget on Home Screen of the iOS device to access website links~\cite{web_clips}. An example is shown in Figure~\ref{fig:web_clip_example}. The red box in Figure~\ref{fig:widget} presents a concrete example of the added widget, and after tapping the icon of this widget, we will access to the webpage shown in Figure~\ref{fig:widget_view}. 

A generalized and simplified Web Clip configuration profile is depicted as Listing~\ref{lst:web_clip}.  
As Listing~\ref{lst:web_clip} depicted, the configuration profile is essentially an XML file consists of key-value pairs. To configure Web Clips, developers can specify value of `Label' (line 10-11) to set the name of the widget, set content of `Icon' (line 8) to designate the icon of the widget displayed on Home Screen. An import key to note is the value of `URL' (line 13-14), which is used to specify the URL for accessing when users tap the widget~\cite{web_clips}. In essence, the Web Clips are just wrappers of remote resource.

To ensure the integrity and remove the warning message of `Unverified Profile' when deploying the Web Clips, developers can opt to sign the Web Clips either by using the certificate issued by Apple or by using a valid SSL certificate issued by a certificate authority. If a Web Clip is signed, the original Web Clip (Listing~\ref{lst:web_clip}) and the certificate are enveloped together to form a new one. The developer within a certificate issued by or the domain within an SSL certificate is an import identity of Web Clips.

\lstset{language=XML}
\begin{lstlisting}[
caption=Snippet setting of Web Clips, 
label=lst:web_clip, 
numbers=left,
frame=tb,
showstringspaces=false,
escapeinside={(*@}{@*)},
basicstyle={\tiny\ttfamily}
]
<?xml version="1.0" encoding="UTF-8"?>
<plist version="1.0">
<dict>
  <key>PayloadContent</key>
  <array>
    <dict>
      <key>FullScreen</key><true/>
      <key>Icon</key><data>...</data>
      <key>IsRemovable</key><true/>
      <key>Label</key>
      <string>Ding Sheng</string>
      ...
      <key>URL</key>
      <string>http://login.tdo-sihgf9-ajfca666.ntum27.cn/#/Login?roomCode=10001</string>
    </dict>
  </array>
</dict>
</plist>            
\end{lstlisting}

\subsection{Dataset and Ethical Considerations}
It's hard to collect Web Clips for there is no centralized distribution channel like iTunes App Store. 
Thanks to the cyber security complaint center of Anonymized Authority\footnote[2]{Anonymized for double-blind review.},
an outlet for receiving complaints of cyber crimes, we get the documents of cyber crimes filed by the authority agent based on the narration of the victim. The non-structured document logging incident description according to per case, including the social media tool the perpetrator used to interact with the victim, the URL provided by the perpetrator to download the Web Clips. We have obtained 1,800 user complaint documents from April 1st to May 11th, 2021 as the corpus for studying the distribution of the Web Clips, and 1,173 unique Web Clips instances for the rest of the study (different cases may refer to the same Web Clips such result in duplicate instances).
For the sake of ethical considerations, these documents are processed by the complaint center to anonymize personal private information as to avoid user privacy violation. 

\subsection{Web Clips Ecosystem Overview} \label{sec:eco}
\begin{figure}
    \centering
    \includegraphics[width=.5\textwidth]{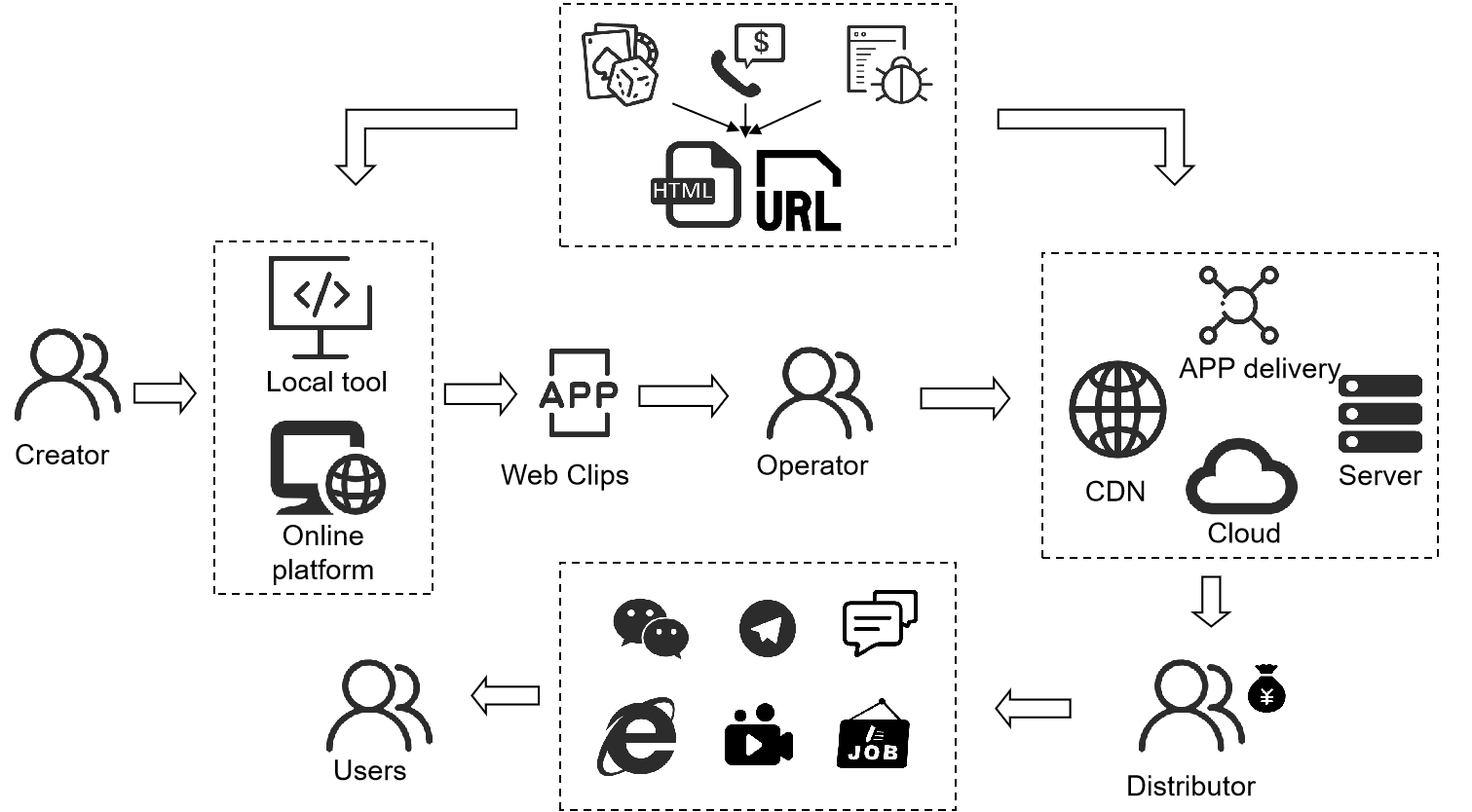}
    \caption{Ecosystem of Abused Web Clips}
    \label{fig:eco}
\end{figure}

In this section, we present the composition of the ecosystem of the abused Web Clips. The ecosystem includes three main participants \ie creator, distributor and operator. Different participants may either come from the same group or belong to different forces that cooperate with each other. The participants and workflow are shown in Figure~\ref{fig:eco}. 

\textbf{Creators:} firstly, creators utilize local tools or online generators to create Web Clips. Target website need to be built beforehand. The corresponding URL of that website is used as input during the Web Clips generation process. 

\textbf{Operators:} after the creation of Web Clips, operators may build Web Clips downloading webpages through subscribing cloud services or outsourcing the APP delivery services. Sometimes, CDN (content distribution network) service is also selected in order to improve user experience. Operator is the key role of the ecosystem.

\textbf{Distributors:} distributors begin to spread the Web Clips download webpages as widely as possible through various medium such as instant messenger, live streaming, social networks, and \etc. Verbal trick is usually designed for attracting users to download and install illicit APPs. A common scenario is that distributors promote online gambling under the guise of high-rewarding part-time jobs. 

Distributors are usually hired by the operators. That is because operators need more workforce to distribute Web Clips in order to attract more people. The division and cooperation among different roles of underground economy are more mature these days. The participants cooperate with but being loosely connected with others, which brings more difficulties for investigation and monitoring.

\section{Creator of the illicit Web Clips}\label{sec:creator}

Given that the Web Clips can be signed either by using certificate issued by Apple or SSL certificate, and such certificate can be used to correlate with an identity of an Apple developer or a domain name, we inspect certificate within the Web Clips in this section. By taking the certificate, Apple can identify the illicit Web Clips and evict them from the iOS host. Moreover, for these Web Clips signed by SSL certificate, we conduct further investigation on the counterpart (namely online generator) of the certificate for law enforcement awareness.
 

\subsection{Creator of the Web Clips}

A signed Web Clips follows the format of \emph{Cryptographic Message Syntax}~\cite{kaliski1998pkcs, ConfigurationProfileReference}, the certificate within the Web Clips contains information of I) `subject'  which indicates the owner (an Apple developer or a domain) of the certificate, II) `issuer' of the certificate in a chain, and the root of the chain can be used to distinguish who (Apple or other certification authority) issued this certificate. The root of the `issuer' can be used to tell the Apple developer apart from domain.

In order to study the creator of the Web Clips, we use the tool of \emph{openssl} to parse the certificate enclosing in the Web Clips. Of the 1,173 unique Web Clips instances, we find 401 Web Clips instances are not signed, and the rest 772, which takes 65.8\% of the overall instances, are signed. Among the signed Web Clips, we are surprisingly to find 724 instances, which take 93.8\% of the signed instances, are signed by using SSL certificate, and the remainder 48 (6.2\% of the signed Web Clips dataset) instances are signed by using certificate issued by Apple however. Note that the 724 instances are signed by 88 unique SSL certificates. The developer and domain identities (IDs) of the top 10 commonly used certificates are listed in Table~\ref{tab:top10_developer}. 

\begin{table}[h]\scriptsize
\begin{threeparttable}
\caption{The identities of the top 10 commonly used certificate}
\label{tab:top10_developer}
\centering
\begin{tabular}{p{4.5cm} p{0.2cm} p{1.8cm}  p{0.2cm}}
\toprule
Top 10 Apple developer's IDs &  Count & Top 10 domain IDs & Count \\
\midrule
iPhone Distribution: peng sun (C379A9YGPS) & 6 & xz.xianke315.com & 134 \\
iPhone Distribution: Chengdu * (95JZQH2Q8R) & 3 & app.htmlpacker.com  & 117 \\
iPCU CA 34128794-d96d-4227-84b8-9d7c8918e683 & 3 & xv8.0mwe.space & 116 \\
iPCU CA e79d51e4-0007-432e-9db7-db13e8864e6f & 2 & iosappnet.com & 70 \\
iPhone Distribution: Weifu He (CLUXRHCPV2) & 2 & yh27.cn & 41 \\
iPCU CA c29cc060-5fd9-48ab-a0ad-6dadd2f054d3 & 2 & 360.ahlpb006.xyz & 35 \\
Apple Development: Chapelle * (HQ3BQ95B6A) & 2 & apk.pingzdoua.xyz & 30 \\
iPhone Distribution: XiaoWei * (3CMSCZUW46) & 1 & xb2.zxwerb08.xyz & 16 \\
Apple Development: lu xu (9K7U3U5ZK5) & 1 & huizhutz.com & 13 \\
iPCU CA 4ab0edcc-ce74-4ca5-94e3-51531060793d & 1 & dsew.ml & 10 \\
\bottomrule
\end{tabular}
\begin{tablenotes}
\item[1] The overlong identities are trimmed by `*' for the space limitation.
\end{tablenotes}
\end{threeparttable}
\end{table}

By comparing the 2nd and 4th columns of this table, we find the creators of the illicit Web Clips are preferred to use SSL certificate rather than certificate issued by Apple to sign their Web Clips instances. The underlying reason is that the illicit Web Clips can't be identified by merely using certificate if the certificate is used by benign one in the mean time. In addition, a SSL certificate is easier to be obtained compared with a certificate issued by Apple. 
Data of the 4th column diverse greatly, which indicate a limited group of SSL certificates is utilized by the Web Clips creators.

Some of the SSL certificates are related to online Web Clips generators. Thus, we extend our study to the Web Clips online generator for a broader understanding of the threat.





\subsection{Online Web Clips Generator}

It is a convention of the online Web Clips generator to sign a Web Clips instance by using the SSL certificate of itself. 
Such that we can access the online Web Clips generator by using the domain name within a SSL certificate. In order to expand the collection of the online generator, we firstly design a crawler to get the first page of the web service reside on the recognized domain name (the 3rd column of table~\ref{tab:top10_developer}), title and body of the HTML file are preserved for later on process. Then we use word segmentation and TF/IDF (Term Frequency-Inverse Document Frequency) algorithm to withdraw keyword of these pages. 21 keywords,  `BetaAPPHost' for example, are obtained in this stage. At last, we use the cyberspace search engine to find the web pages that match these keywords. Eventually, we successfully expand the domain collection to 955 for our study, which is about ten folds of the former 88 domains.

Even though under most cases online Web Clips generators are not actively misused, understanding the high level network infrastructure information of them provides valuable clues for law enforcement. A knowledge base of online Web Clips generators could be built and maintained in order to understand the location distribution, underlying infrastructure, domain name changing records and etc. Under some cyber-investigation cases, if illicit Web Clips are found to be related with a certain online generator, the knowledge base could be utilized to get more clue.

Therefore, in the rest of this section, we study the IP location distribution, the infrastructure provider, and domain name resolution of the online Web Clips generators.



\textbf{IP location:} The following figure~\ref{fig:IP loc} shows the IP location distribution of online Web Clips generator domain names. Since we use key words of generators in Chinese as study objects, the IP addresses of about half generators are located in China mainland, while others are located in HK China, US, Korea, Spain and Philippines. 

\begin{figure}[h]
    \centering
    \includegraphics[width=.45\textwidth]{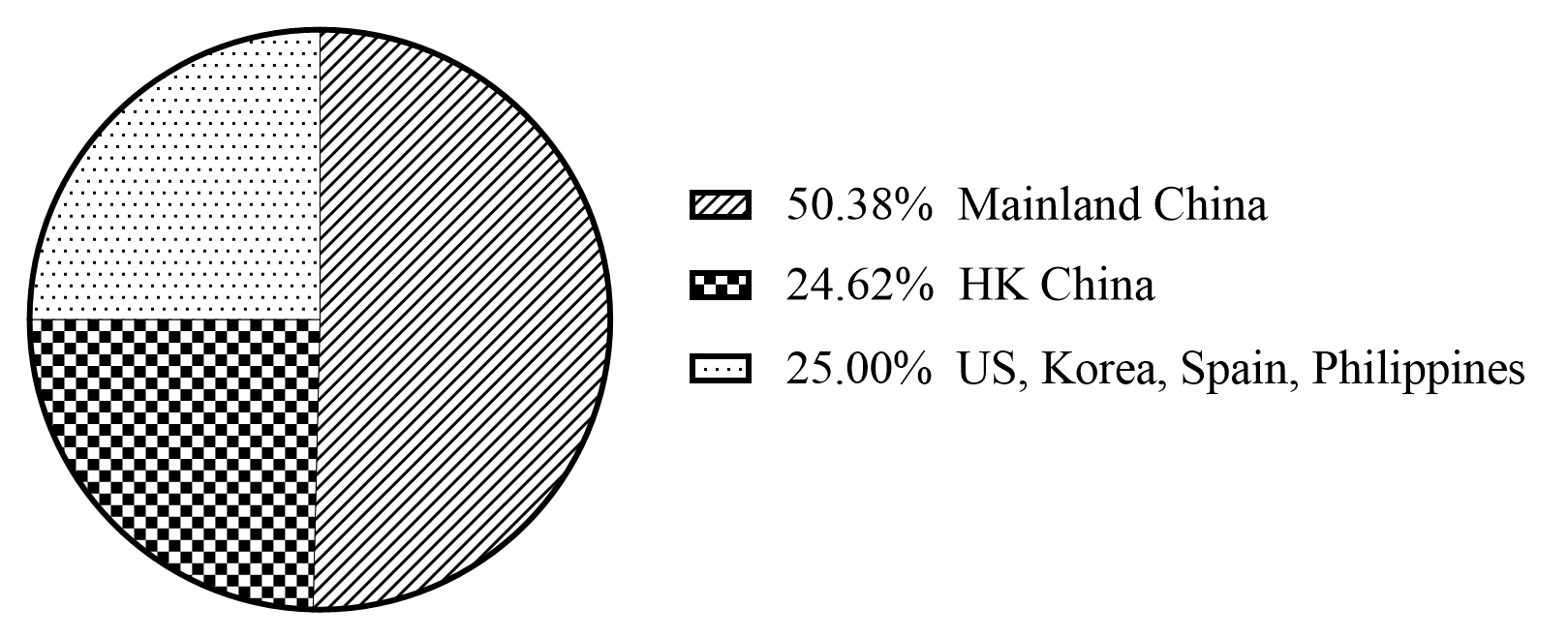}
    \caption{IP location distribution}
    \label{fig:IP loc}
\end{figure}

\textbf{Infrastructure:} Online Web Clips generators are websites deployed on some kind of infrastructure. About 49.8\%  online Web Clips generators provide services through public cloud platform such as Alibaba cloud, Tecent cloud, Amazon cloud and \etc. About 18\%  generators provide services through traditional network operators. The left generators do not show deployment information. What should be mentioned is that about 42\% domain names of the generators have been filed in official authorities. Some filed information of domain names is irrelevant with the online generator, for example a domain name is filed by a food supply corporation. The main reason is that the transaction of filed domain names is quite popular and easily being abused in underground economy nowadays. 

\begin{figure}[h]
    \centering
    \includegraphics[width=.48\textwidth]{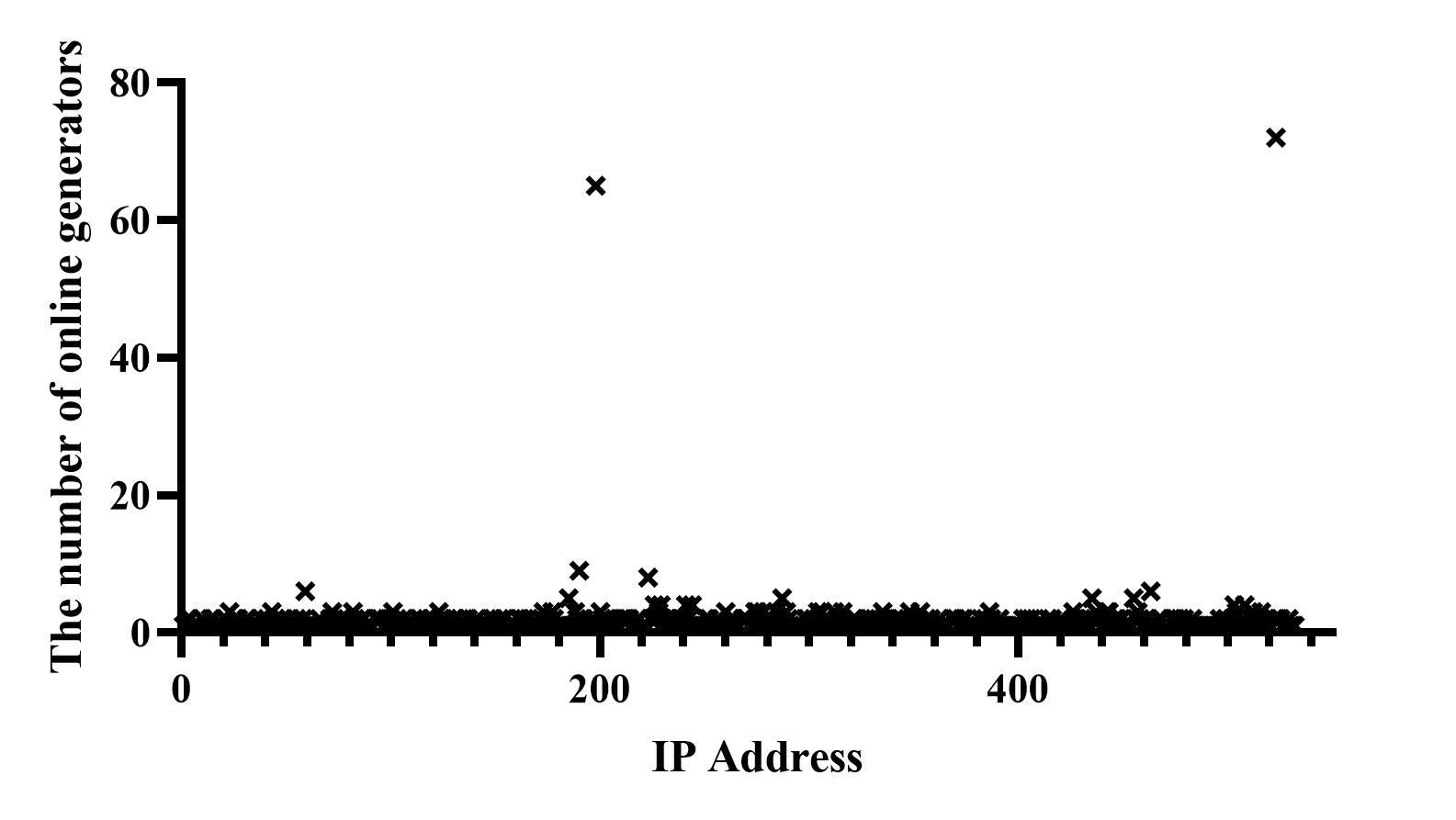}
    \caption{Domain name aggregation of online Web Clips generators}
    \label{fig:ds agg}
\end{figure}

\textbf{Domain name aggregation:} We find that some domain names of online Web Clips generators are resolved to the same IP. Figure~\ref{fig:ds agg} shows that all the 955 domain names correspond to 512 IP addresses. Generally, an IP address is used by less than 5 domain names. In special cases, one IP address is used by more than 70 domain names, and all the domain names correspond to the same online Web Clips generator website. This kind of website might use many domain names for some special reason like anti-blocking, which needs to be payed more attention on. 

\section{Purpose of the illicit Web Clips} \label{sec:purpose}

This section seeks to find out the purpose the illicit Web Clips built for. To accomplish this task, we develop the taxonomy of the Web Clips, then triage the 1,173 illicit Web Clips instances. 

\subsection{Taxonomy of Web Clips} \label{sec:web_clip_taxonomy}

Taxonomy of Web Clips is developed in terms of the content it delivers because the Web Clips is a carrier of web resource. 
The development starts with the initial categories of IC3~\cite{ic3_report}, then the categories are refined iteratively according to the ongoing triage process. To ensure the accuracy, domain experts are involved to supplement/remove the categories and set the summary of the category. Laws on online gambling are different from country to country. In some EU countries online gambling is legal, but there are still many countries like China that prohibit online gambling. Therefore, illegal online websites in this paper are those containing content that is not allowed to be distributed by certain countries. The final taxonomy consists of 6 categories: `Fraud', `Pornography', `Gambling', `Copyright', `Others', and `Unknown'. 
The category is further divided into sub-category, the category `Fraud' is separated into `Fake Loan', `Investment', `Crowdsourcing scams', `Digital Currency', `Fake Transactions', for example. Summary of each sub-category is depicted in the third column of Table~\ref{tab:taxonomy} in the appendix.

\subsection{Triage the Web Clips} \label{sec:web_clip_triage}

The 1,173 illicit Web Clips are manually triaged by co-authors of this paper with a thorough understanding of the taxonomy. Note that 1) One Web Clips may fall into multiple categories, we take in the distinct visual element as the criteria for categorizing. \eg, pornography in a Web Clips is actually designed for attracting victim to visit a gambling page, we categorize the the Web Clips as `Pornography', for the tiny window on the right-bottom corner takes a small portion of the view. 2) There is intersection between the categories. \eg, the dishonest gambling can be categorized as both `Fraud' and `Gambling' categories, for lacking context of the Web Clips, we categorize it as `Gambling' nonetheless. 3) All minor categories are categorized as `Others'.

\begin{figure}[h]
    \centering
    \includegraphics[width=.40\textwidth]{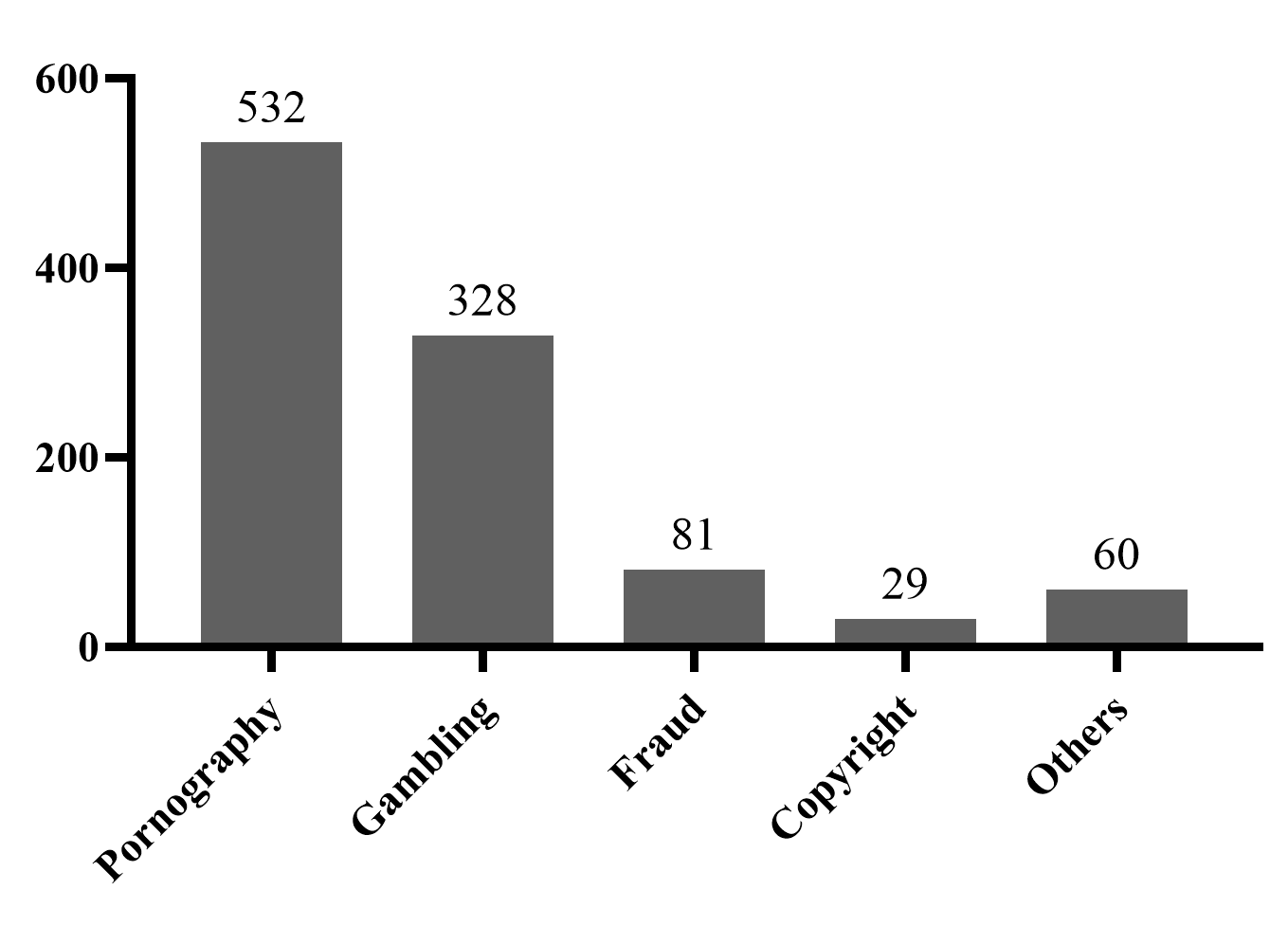}
    \caption{Web Clips triage result.}
    \label{fig:taxonomy}
\end{figure}

Triage result in figure~\ref{fig:taxonomy} shows the Web Clips are build for purpose of `Pornography', `Fraud', and `Gambling', which takes 45.35\%, 27.96\% and 6.91\% of the dataset separately, with a total 80.22\% of the entire dataset. If a website link contained in Web Clips is out-of-date and the type cannot be clearly identified, then the Web Clips will be marked as `Unknown' type. The `Unknown' type is not shown in the figure.

\begin{figure}[h]
    \centering
    \includegraphics[width=.5\textwidth]{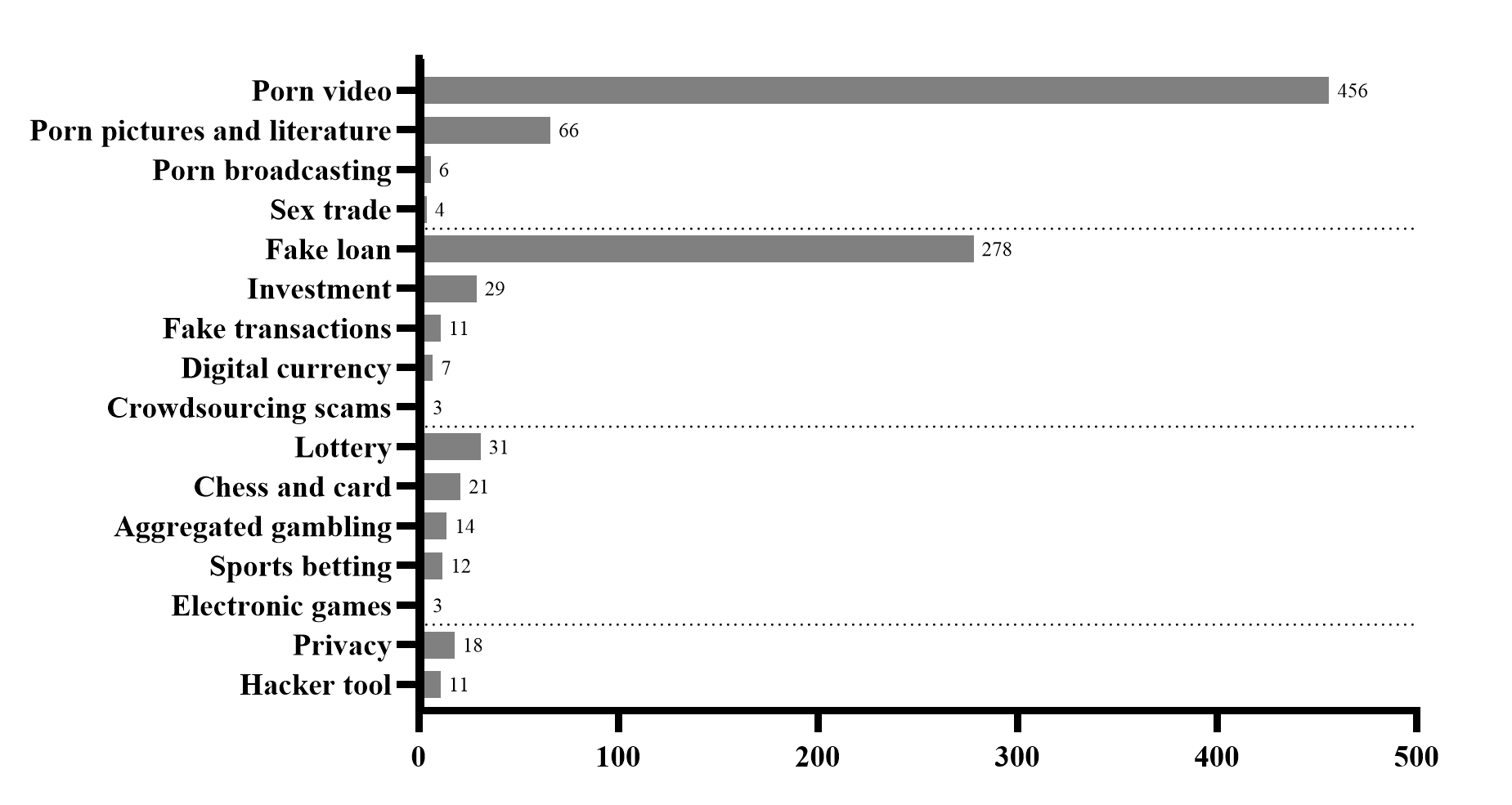}
    \caption{Web Clips sub-category triage result.}
    \label{fig:subcategory}
\end{figure}

Triage result of sub-categories in figure~\ref{fig:subcategory} reveals: 1) `Fake loan' takes the prominent portion of the `Fraud' category, indicating such scheme is widely used by fraudsters to seize money from victim; 2) `Pornographic video' is the major sub-category of the `Pornography' category, indicting perpetrators prefer to distribute video rather than others; 3) The small variance of the sub-categories of `Gambling' indicts there is no obvious preferable schemes for perpetrators. Note that there is bias results from the `Unknown' category, however such category takes small portion of the dataset so that can be ignored.

The triage result of the illicit Web Clips also reveals the types of prevalent cyber crimes in current days. Breakdown of the triage result is depicted in Table~\ref{tab:taxonomy} in the appendix.




\section{Distributor of the illicit Web Clips}\label{sec:distributor}

\begin{figure}[h]
    \centering
    \includegraphics[width=.45\textwidth]{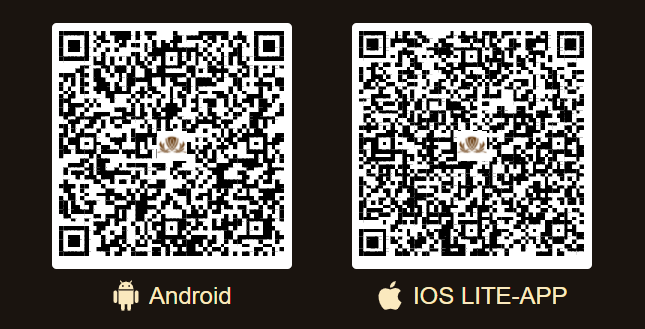}
    \caption{Web Clip download information}
    \label{fig:QR}
\end{figure}

As mentioned before, Web Clips are abused for launching illicit URLs as APPs in iOS. Then, we study how the Web Clips are distributed. Usually illicit APPs are provided by download webpages. A typical download webpage is shown in figure~\ref{fig:QR}, which offers QR code for Android APK or iOS light APP (Web Clips). 
There are various ways to popularize the download webpages. As mentioned before, we collect user compliant messages during the year of 2021. About 1800 illicit APPs (Web Clips for iOS) and their distribution processes are described . 
Event extraction (EE) method to recognize events of illicit APPs distribution from user complaint messages automatically. EE is an advanced form of Information Extraction (IE), which focuses on gathering knowledge about incidents found in texts and automatically identifying information about what happened. An EE task is composed of event trigger, event type, event schema, event augment and augment role.

An EE task can be divided into three sub-tasks including event detection, event augment recognition and augment role recognition. The following figure~\ref{fig:EE method} depicts EE task for distribution of illicit APPs and gives an example. The basic event schema is designed as ``A \textbf{suspect} utilizes \textbf{verbal trick} to distribute an \textbf{object APP} to \textbf{victims} via a \textbf{medium}''. A complaint message (``A stranger recommended online part-time job through WeChat group chat, in fact users are lured to download gambling APP'') is used as an example. Firstly, during the event detection stage, an event is triggered by `recommending' and the event type is `Distribution'. Next, event augment (EA) recognition is performed to find entities such as part-time job, WeChat, gambling APP, and etc. These entities are learned from sample data and provided as prior knowledge for EA recognition. At last, augment roles such as `subject', `verbal trick', `victims' and `medium' are recognized. 
\begin{figure*}[h]
    \centering
    \includegraphics[width=1\textwidth]{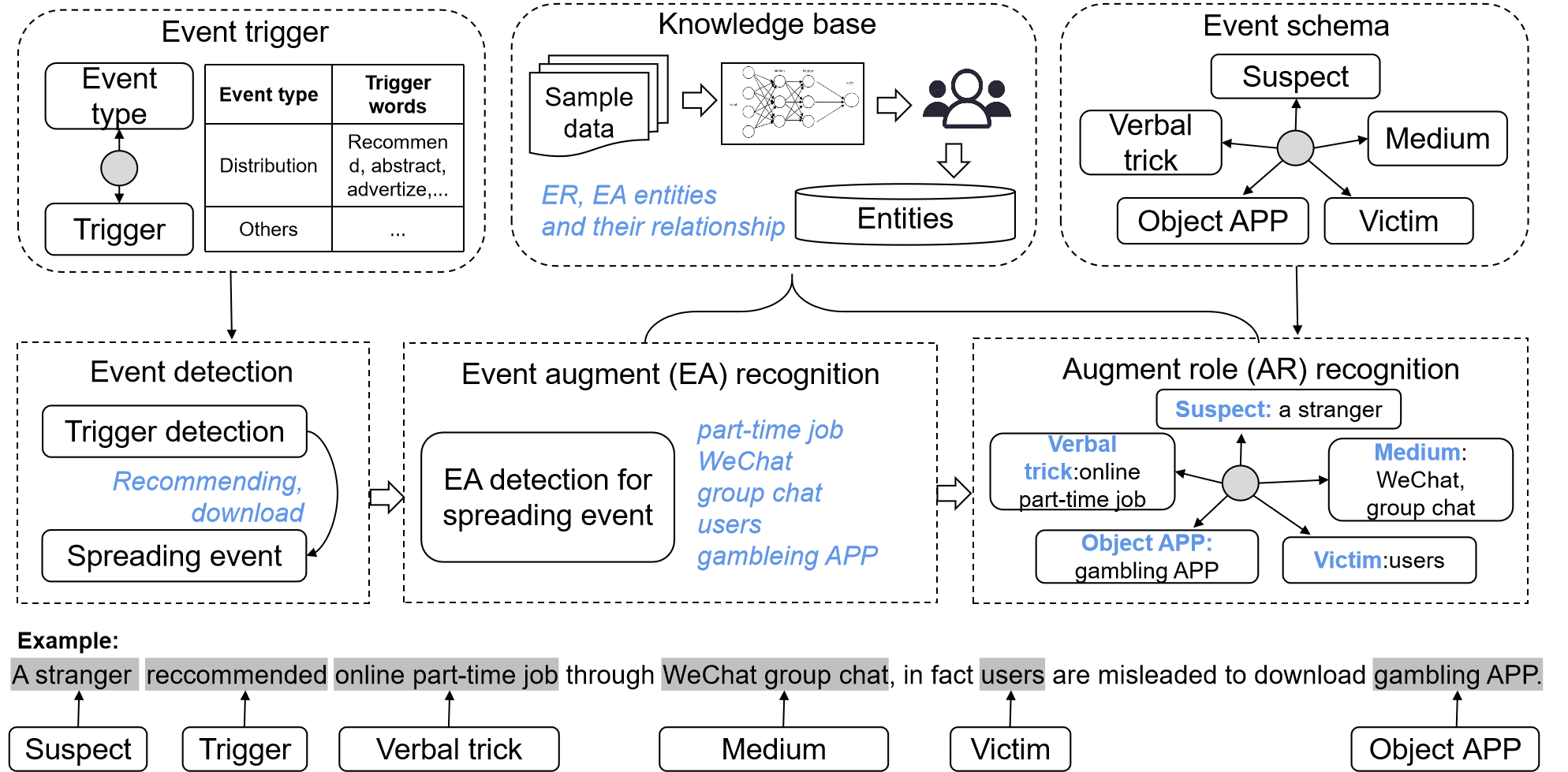}
    \setlength{\belowcaptionskip}{-3ex}
    \caption{Event extraction task}
    \label{fig:EE method}
\end{figure*}

Through the above EE analysis, we recognize about 711 distribution events from the 1800 complaint messages. About 42$\%$ distribution events are related to gambling APPs which account for the highest proportion. And 31$\%$ distribution events spread pornography APPs. This may be because gambling advertising activities are more complex and well-designed. Therefore, more distribution events can be detected.

\begin{figure*}[h]
    \centering
    \includegraphics[width=1\textwidth]{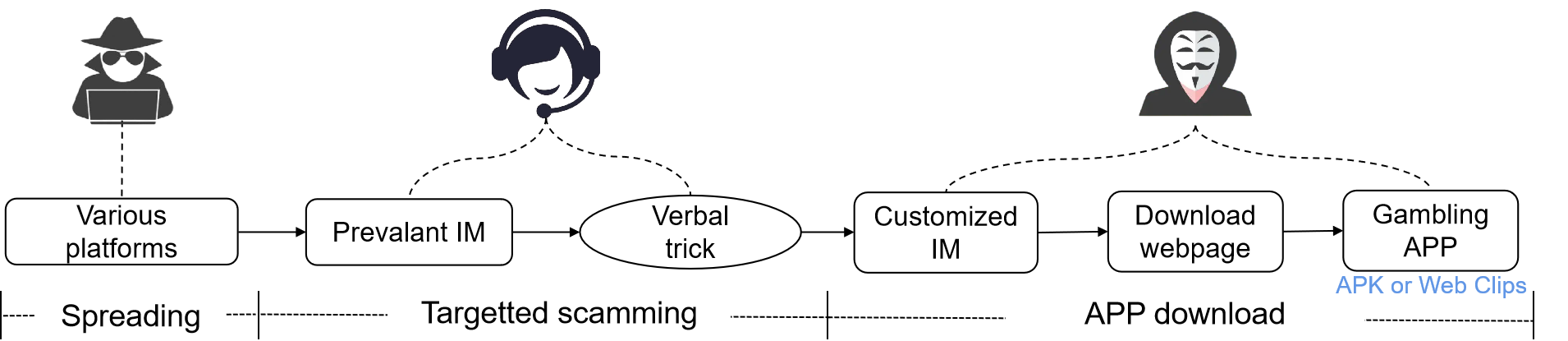}
    \setlength{\belowcaptionskip}{-3ex}
    \caption{Typical distribution process of gambling APPs}
    \label{fig:gambling APP}
\end{figure*}

\begin{figure}[h]
    \centering
    \includegraphics[width=.48\textwidth]{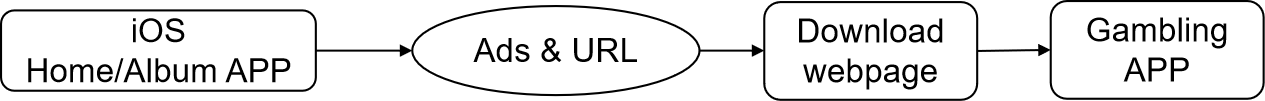}
    \caption{A popular distribution way of illicit APPs on iOS}
    \label{fig:ios spread}
\end{figure}

\begin{figure}[h]
    \centering
    \includegraphics[width=.48\textwidth]{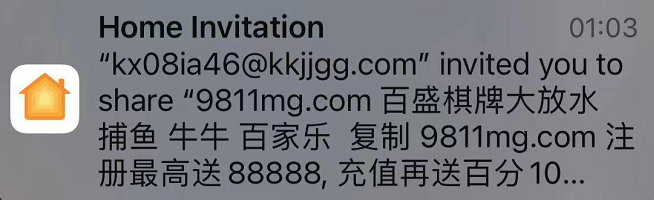}
    \caption{An example of a new distribution method on iOS. The text in the figure means that an online gambling website contains multiple gambling games, and claims that after copying the domain name to visit the website, users will receive a large amount of fees after registering or making deposits.}
    \label{fig:iOS example}
\end{figure}

From our analysis, the distribution process of gambling APPs are shown in figure~\ref{fig:gambling APP}. Firstly, advertisements are posted on various platforms, such as Internet forum, social network, recruiting website, live streaming platform and etc. We call the activity as `spreading stage' which aims for attracting more people. The second stage is `targeted scamming' during which verbal trick is organized in order to scam a particular victim. Some prevalent instant messengers (IM) such as WeChat, Telegram are often adopted to interact with victims. And the last stage of distribution is `APP download'. Victims are guided to download and use the gambling APPs. Usually, a customized IM is used to interact with victims because they are more difficult to be detected and investigated. Our analysis on the complaint data shows that more than 20 customized IMs are mentioned.
In addition, various platforms and prevalent IMs that are used for distributing illicit APPs are depicted by the following figure~\ref{fig:platform}. we can draw some conclusions based on the results. IM, live streaming and group chatting are the most commonly used medium for illicit APPs distribution. What should be explained is that IM and group chatting may have some overlap. Here, IM means one-on-one conversation while group chatting refers to the interaction in any kind of chatting groups. There are various platformed used for distributing gambling APPs, IMs are mostly relied on. Live streaming is mainly used for distributing pornography APPs. 
\begin{figure}[h]
    \centering
    \includegraphics[width=.48\textwidth]{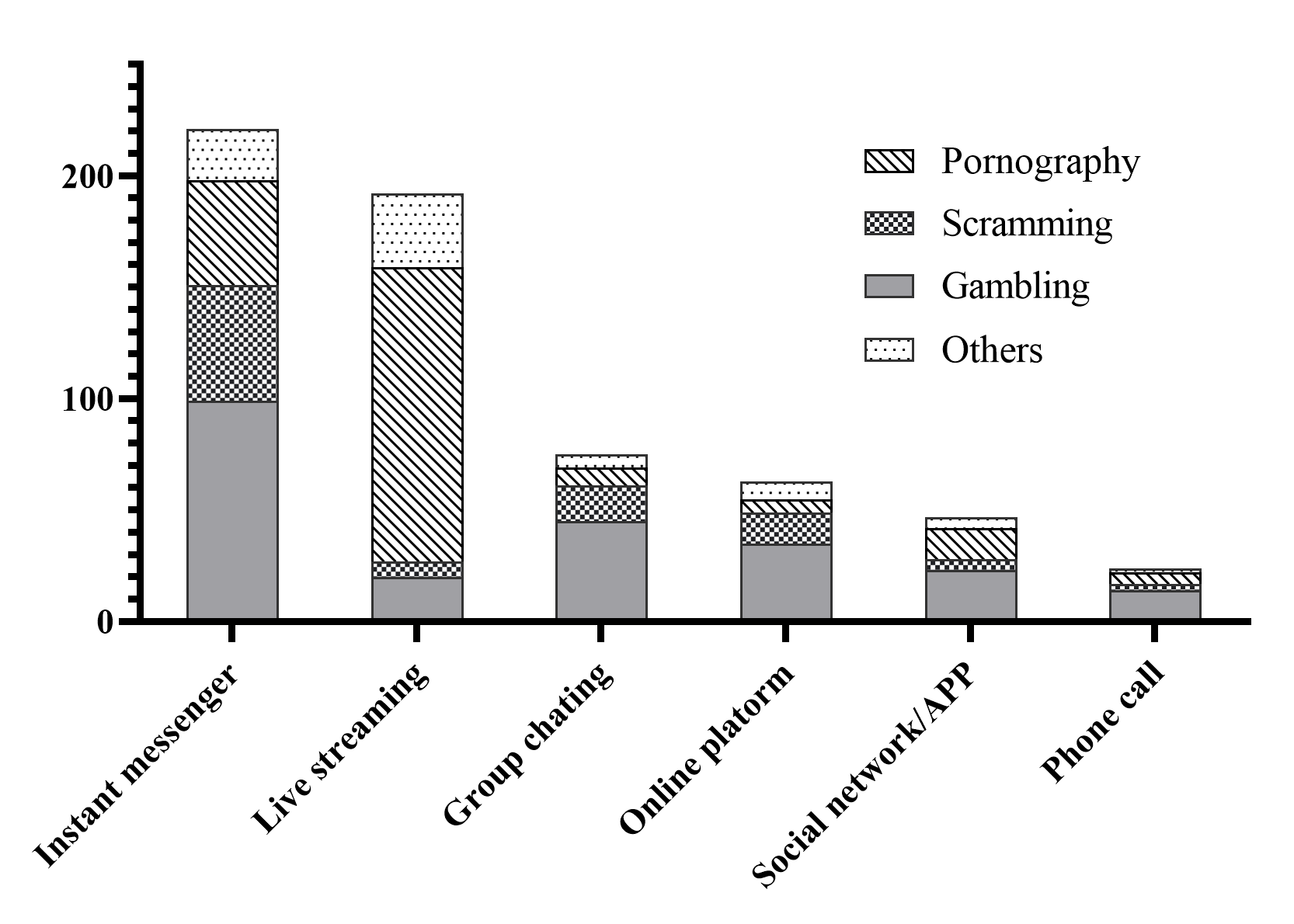}
    \caption{Platforms used for distributing illicit APPs}
    \label{fig:platform}
\end{figure}
We also find a relatively new distribution way of illicit APPs on iOS , which is shown in figure~\ref{fig:ios spread}. In recent years, many personal emails are collected and sold unlawfully. Then, criminals send advertising texts and illicit URLs through iOS Home share, album or calendar functions using the large amount of emails  (relying on the iMessage mechanism). If an email address is registered an Apple ID, iOS devices logged in by that Apple ID will receive the illicit advertising texts. An example is given in Figure~\ref{fig:iOS example}. 

\section{Operator of the illicit Web Clips}  \label{sec:operator}
Since most content of the Web Clips originate from the back-end server, in this section, we profile the Web Clips operators by means of 1) inspecting the network property they owns for facilitating the illicit activity; 2) investigating technical tricks they use for bypassing supervision. The study conduct on the counterpart server of the 1,173 illicit Web Clips.

\subsection{Network Property of the Operator}
\label{sec:Dist_res}

By equipping with adequate network resource, operator remains at the forefront of the competition with the law enforcement. Inspired by works of~\cite{gao2021demystifying, yang2019casino}, we inspect domain name property, IP property, and web resource property of the illicit Web Clips operator. 

\subsubsection{Domain Name Property}

WHOIS record of a domain name is an important fingerprint of the operator, these information, including registrar and registrant email address, give hints for law enforcement to take action against the illicit activities. By applying the \emph{python-whois} library against the domain names specified by the 1,173 illicit Web Clips, we get 65 registrars and 84 registrant email addresses of these domains. The top 10 most frequently used registrar are depicted in Table~\ref{tab:top10rege}. Note the the register's email addresses are omitted herein for they are fully anonymized for public access according to the GDPR (General Data Protection Regulation), \eg, `abuse@godaddy.com' for GoDaddy.com, LLC., and `abuse@support.gandi.net' for GANDI SAS.

\begin{table}[h]
\begin{threeparttable}
\caption{Top 10 Registrar used in criminal Web Clips}
\label{tab:top10rege}
\begin{tabular}{lcc}
\toprule
Registrar & Count & Percent (\%) \\
\midrule
GoDaddy.com, LLC   & 277 & 24.01  \\
GANDI SAS     & 257 & 22.27   \\
Gname.com Pte. Ltd.  & 123  & 10.65   \\
NameSilo, LLC  & 76  & 6.58  \\
DYNADOT LLC  & 66  & 5.71   \\
GMO INTERNET, INC.  & 40  & 3.46    \\
Alibaba Cloud Computing Ltd.  & 39  & 3.37   \\
Chengdu west dimension * Co., LTD & 29  & 2.51    \\
Alibaba Cloud Computing * Co., Ltd.   & 25  & 2.16    \\
Xin Net Technology Corporation   & 14  & 1.21   \\
\midrule
\textbf{Total} & \textbf{896} & \textbf{81.93} \\
\bottomrule
\end{tabular}
\begin{tablenotes}
\item[1] The overlong registrars are trimmed by `*' for the space limitation.
\end{tablenotes}
\end{threeparttable}
\end{table}

Although the register's email addresses are anonymized, law enforcement can resort to registrar to look up such information without violating GDPR. The cooperation between different sectors drives operator to get domain name from the oversea registrar (relative to China), such that the top six oversea registrars are used by 72.68\% of the entire samples, which raise the difficulty of crime investigation. Luckily, there are four remainder domestic registrars, which takes 9.25\% of the entire Web Clips instances. 

In addition, we make a statistics on the TLDs (Top Level Domains) of the domain name designated by the illicit Web Clips. The statistical result is depicted in Table~\ref{tab:top10tld}. 

\begin{table}[h]
\caption{Top 10 TLDs used in criminal Web Clips}
\label{tab:top10tld}
\begin{tabular}{lccc}
\toprule
TLD & Number (\#) & New gTLD & Percent (\%) \\
\midrule
xyz     & 365 & $\checkmark$ &  31.15 \\
com     & 339 &  &  28.94 \\
today      & 68  &$\checkmark$&  5.78 \\
cn     & 62  &       &  5.29  \\
ltd   & 52  &$\checkmark$&  4.43  \\
world     & 45  &$\checkmark$&   3.81  \\
top      & 42  &$\checkmark$& 3.57 \\
vip    & 39  &$\checkmark$&  3.32  \\
IpAddr     & 26  &$\checkmark$&  2.21  \\
cc     & 22   &$\checkmark$&   1.84 \\
\midrule
\textbf{Total}   & \textbf{1060} &       &  \textbf{90.34} \\
\bottomrule
\end{tabular}
\end{table}

It is counterintuitive that the scarce TLD `com', which takes 28.94\% in total, are wildly used by operator, later investigation reveals that such domain name are mainly used by Web Clips of `Gambling' category, which is legal in some regions or countries to be used directly. Note that the easy-to-acquire \emph{New gTLD} (3rd column of Table~\ref{tab:top10tld}) contributes a large portion of the misused TLDs, suggesting firm surveillance should be deployed on them, such that the illicit domain name can be identified in time.

\subsubsection{IP Property}
The flexible mapping between domain name and IP address allows a domain name be re-resolved to another IP address and vice versa. In order to make clear the intricate relationship among these domains (1,173), we firstly correlate the domains by using the IP address if they are resolved to a same IP address, then we study the cloud service provider the operator favorite to use.

\begin{figure}[h]
    \centering
    \includegraphics[width=.46\textwidth]{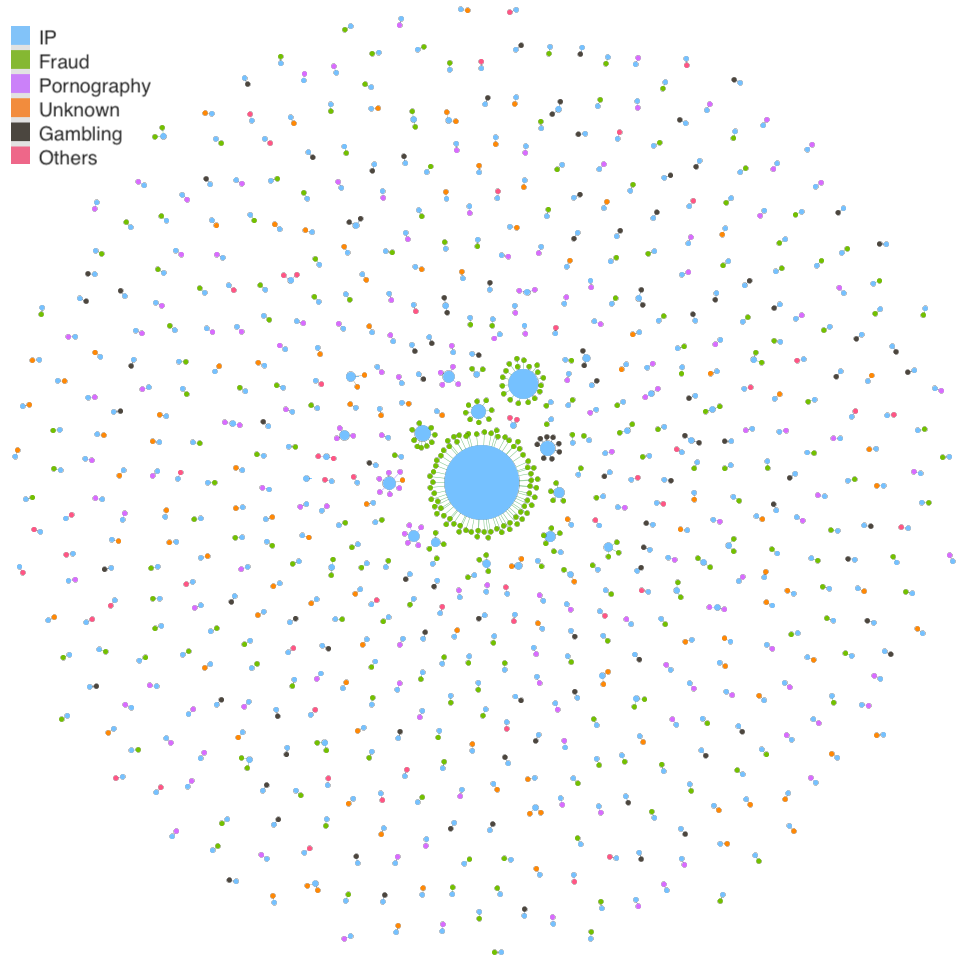}
    \caption{Correlate domain name by using IP address.}
    \label{fig:network}
\end{figure}

The 1,173 domain names are eventually resolved to 606 IP addresses, the relationship among these domain names is depicted in Figure~\ref{fig:network}. Different node color in this figure represents different categories (cf. \S~\ref{sec:purpose}) this domain name belongs to, except the blue node with variant size represents the IP address for clustering. We highlight that the domain names of `Pornography' and `Fraud' categories are clustered by limited IP address, suggesting the back-end servers are managed by limited operators. For example, the largest cluster in Figure~\ref{fig:network} consists 63 domain names (\eg, `mhb01.*.xyz', `thb05.*.xyz') for conducting `Fake loan' scheme, all of them are correlated by IP address of `52.77.67.*'. 

\begin{figure}
    \centering
    \includegraphics[width=.46\textwidth]{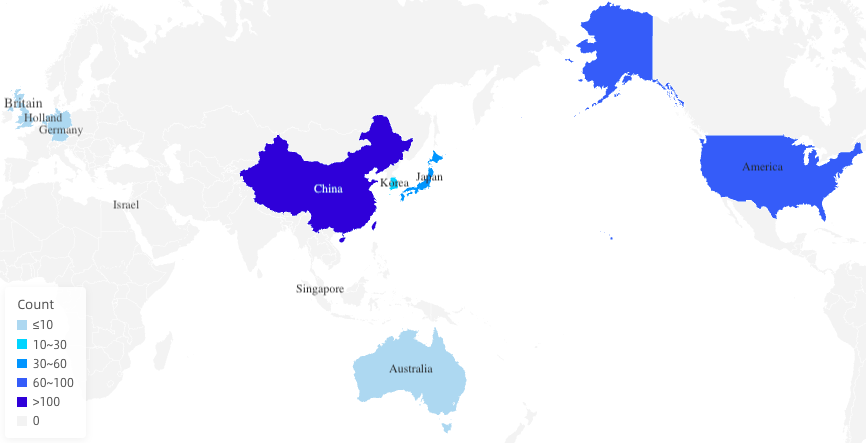}
    \caption{Location of the IP address.}
    \label{fig:map}
\end{figure}

We further mark the 606 IP locations on a world map, which is depicted in figure~\ref{fig:map}. As note, 132 IP addresses with a 22\% of the entire dataset, locate on oversea zone or country of China. This is the underlying reason why the operator provides continuous web service on the above mentioned IP address (\eg, `52.77.67.*') regardless of the regulation on the domain names (\eg, `mhb01.*.xyz', `thb05.*.xyz').

\subsubsection{Web Resource Property}

Operators are free to copy, deploy, and publish the web resource for delivering the illicit or forged content to victims.
The ownership of the web resource discloses the operators under the hood. In order to narrow the large amount of IPs down to limited operators, we cluster the web resource of these IPs by analyzing the similarity of them. Since the retrievable resource is the html file, we analyze the similarity of these files.

\begin{figure}[h]
    \centering
    \includegraphics[width=.46\textwidth]{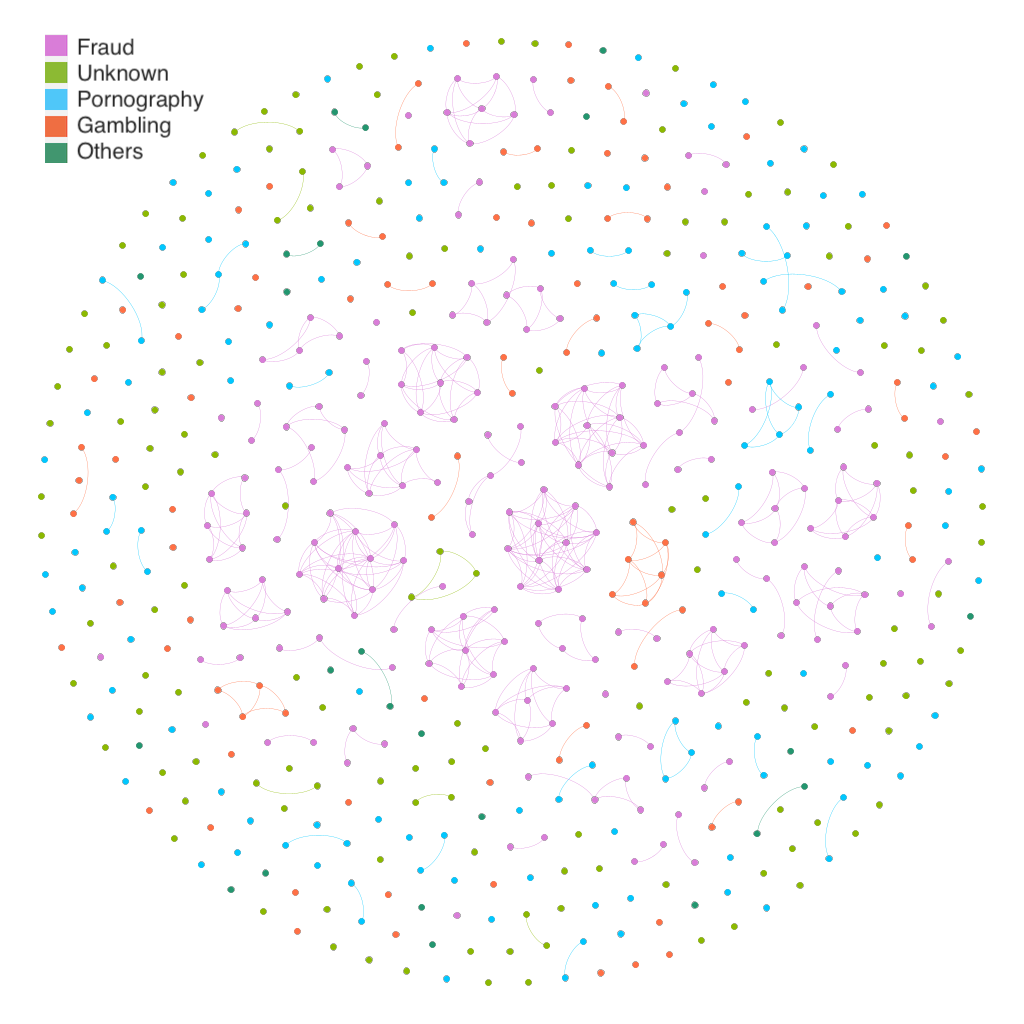}
    \caption{Clustering result by using similarity analysis.}
    \label{fig:sim_clustering}
\end{figure}

Rather than analyzing the literal similarity, we measure the structural similarity of two html files. Study of Tang~\etal~\cite{tang2021similarity} reveals that the performance of bucketing the tags of an html file~\cite{HTMLSimilarity} beforehand surpasses those compare the tags directly~\cite{page_compare}, we opt to use the former strategy for clustering the html files. Our clustering is built on the python library of \emph{HTMLSimilarity}~\cite{HTMLSimilarity}, which bucketing the tags in an html file to fix-length (512) vector. In addition, to avoid the pair-wise comparison for all html files of two different IPs with the complexity of $O(n^2)$, we only compare the first page of these IPs; To deal with the page rendered by Javascript (\eg, Angular, React, or Vue.js) that contains less tags as features for similarity analysis, we use the python library \emph{Esprima}~\cite{Esprima} to get AST (Abstract Syntax Tree) of the Javascript code, then traverse the obfuscation-resilient tags in the AST and bucketing them into fix-length (512) vector. The similarity of two pages is measured by \emph{Euclidean distance} of the vectors, and two IPs are connected if the distance less than the threshold of 0.01.
The source code is available at Web Site\footnote{Anonymized for double-blind review.}, and the clustering result is depicted as Figure~\ref{fig:sim_clustering}.

By using the similarity analysis, we successfully cluster the 606 IPs to 392 groups. From Figure~\ref{fig:sim_clustering}, we conclude that: 1) multiple hosts deployed with the same `Fraud' category web resource indicates these hosts are managed by the same operator or the web resource is acquired from the same developer; 2) the diversity of web resource of `Pornography' and `Gambling' categories indicates there are a large number of operators involve in such scheme or the same operator owns versatile web resource to conduct the illicit activities. To sum up, compared with `Pornography' and `Gambling' operators, `Fraud' operators are more willing to use templates (\eg, website template) of web resources. 
   

\subsection{Tricks of Anti-Crawler}

We have witnessed the sophisticated technique the operator used to evade detection. Such techniques include: 1) By using shortlink to designate the target domain at will; 2) Deny access without a specific parameter in the URL; 3) Use web page developing technique \emph{iframe} to conceal the illicit domain name~\cite{Fraud_China}. 
In this section, we highlight the anti-crawler techniques used in the 1,173 websites to have deeper insight of the operator.

Considering that the adversary usually utilizes crawler to get resource of the illicit websites~\cite{yang2019casino, gao2021demystifying}, the operator defeats the crawler by checking the run-time environment of the client or by exploiting drawback of the crawler. For example, by checking \emph{User-Agent} field in the request header, the operator refuses the request originate from a Non-iOS devices, we found 92 websites, which takes 7.8\% of the 1,173 websites, are deployed with such scheme. Another prevalent used technique is to redirect the request multiple times by using the acquisition of the domain name/IP property, this setting will drain the resource of a crawler if it follows such redirection. Such scheme is deployed to 504 websites, about 43\% of the dataset.

\section{MITIGATION}\label{sec:mitigation}
In this section, we give some concluded mitigation measures based on the whole paper's results.
\begin{itemize}
    \item Web Clips can be deployed by using Apple Configurator 2, email, webpage or over the air~\cite{ConfiguringDev}. This work studies the Web Clips deployed via website. Of note, such Web Clips is misused in cyber crime. We strongly suggest Apple disable the support for Web Clips deployed through website.
    \item Most websites used for generating online Web Clips also provides other similar services such as APK packaging, TestFlight signing and etc. In order to gain more profit, the websites usually adopt weak user identity authentication and perform few check work on the URL being packaged. Therefore, they are more easily to be abused in illicit activities. More supervision should be placed on the App generation websites.
    \item Considering the new distribution method of illicit messages on iOS devices, for one thing, we need to do more to protect personal information. For another thing, we suggest that rules for sending messages through iOS Home share or Album should be regulated. Messages can't be sent arbitrarily.
    \item The streamline communication and collaboration are desired between the stakeholders in the process of criminal investigation, especially for these cross sectors, regions or countries. Therefore, international cooperation on investigation and enforcement is essential. 
\end{itemize}

\bibliographystyle{ACM-Reference-Format}
\bibliography{ACSAC}


\appendix

\section{Appendix}
\begin{table*}
\caption{Taxonomy and categorize result.}
\label{tab:taxonomy}
\centering
\footnotesize

\resizebox{\linewidth}{!}{
\begin{tabular}{lllll}
\toprule
Category & Sub-category & Description & Number (\#) & Percentage (\%) \\
\midrule
\multirow{7}{*}{Fraud} & Fake loan  & Ask commission for the Loan, \eg, Credit Loans. & 278 & 23.69 \\
\cmidrule(lr){2-5}                  
                  & Investment & Induce victims to make purchases based on false information, \eg, Stock. & 29 & 2.47 \\
\cmidrule(lr){2-5}                  
                  & Crowdsourcing scams & Induce victims to finish task without return. \eg, buy to promote the product. & 3 & 0.26 \\
\cmidrule(lr){2-5}                  
                  & Digital currency & High profits is claimed for the investments on digital currencies. & 7 & 0.60 \\
\cmidrule(lr){2-5}                  
                  & Fake transactions & Goods or services are never received. & 11 & 0.94 \\
\cmidrule(lr){2-5}
                  & \textbf{Total} & - & \textbf{328} & \textbf{27.96} \\
\cmidrule(lr){1-5}

\multirow{5}{*}{Pornography} & Pornographic video & Offer paid service for accessing pornographic videos. & 456 & 38.87 \\
\cmidrule(lr){2-5}                  
                  & Pornographic pictures and literature & Offer paid service for accessing pictures and literature. & 66 &  5.63\\
\cmidrule(lr){2-5}                  
                  & Pornographic broadcasting & Offer paid service for accessing pornographic broadcasting. & 6 & 0.51 \\ 
\cmidrule(lr){2-5}                  
                  & Sex trade & Offer service for sex trade. & 4 & 0.34 \\ 
\cmidrule(lr){2-5}
                  & \textbf{Total} & - & \textbf{532} & \textbf{45.35} \\
\cmidrule(lr){1-5}

\multirow{6}{*}{Gambling} & Chess and card & Gambling built on chess and card, \eg, card, mahjong. & 21 & 1.79 \\
\cmidrule(lr){2-5}                  
                  & Lottery & Provide unauthorized lottery service. & 31 & 2.65 \\
\cmidrule(lr){2-5}                  
                  & Sports betting & Bet on the result of sports. & 12 & 1.02 \\ 
\cmidrule(lr){2-5}                  
                  & Electronic games & Entertainment games for the purpose of gambling. & 3 & 0.26 \\
\cmidrule(lr){2-5}                  
                  & Aggregator Gambling & Aggregate multiple types of gambling platforms in one site. & 14 & 1.19 \\
\cmidrule(lr){2-5}
                  & \textbf{Total} & - & \textbf{81} & \textbf{6.91} \\
\cmidrule(lr){1-5}
                  
\multirow{3}{*}{Copyright} & Hacker tool & Provide code injection tools to modify other app. & 11 & 0.94 \\
\cmidrule(lr){2-5}
                  & Piracy & Provide pirated video or software. & 18 & 1.53 \\
\cmidrule(lr){2-5}
                  & \textbf{Total} & - & \textbf{29} & \textbf{2.47} \\

\cmidrule(lr){1-5} 
Others & - & Minor categories does not fall into the above categories. & 60 & 5.11\\

\cmidrule(lr){1-5} 
Unknown & - & Access denied \eg, Requiring access code/sign-up, or banned/shutdown. & 143 & 12.19\\

\cmidrule(lr){1-5}                  
                  & \textbf{Total} & - & \textbf{1173} &\textbf{100} \\ 
\bottomrule
\end{tabular}}
\end{table*}

\end{document}